Accepted Manuscript. Published version available in IEEE Transactions on Computational Social Systems. DOI: 10.1109/TCSS.2024.3376732.

# Artificial Intelligence can Recognize Whether a Job Applicant is Selling and/or Lying According to Facial Expressions and Head Movements Much More Correctly Than Human Interviewers

Hung-Yue Suen, Kuo-En Hung, Che-Wei Liu, Yu-Sheng Su*, and Han-Chih Fan

***Abstract*—Whether an interviewee's honest and deceptive responses can be detected by the signals of facial expressions in videos has been debated and called to be researched. We developed deep learning models enabled by computer vision to extract the temporal patterns of job applicants' facial expressions and head movements to identify self-reported honest and deceptive impression management (IM) tactics from video frames in real asynchronous video interviews. A 12- to 15-minute video was recorded for each of the N=121 job applicants as they answered five structured behavioral interview questions. Each applicant completed a survey to self-evaluate their trustworthiness on 4 IM measures. Additionally, a field experiment was conducted to compare the concurrent validity associated with self-reported IMs between our modeling and human interviewers. Human interviewers' performance in predicting these IM measures from another subset of 30 videos was obtained by having N=30 human interviewers evaluate three recordings. Our models explained 91% and 84% of the variance in honest and deceptive IMs, respectively, and showed a stronger correlation with self-reported IM scores compared to human interviewers.**

## I. Introduction

IN employment interviews, job applicants aim to create a positive impression to sway the interviewers' evaluations and increase their likelihood of securing a job offer [1]. This strategy is known as impression management (IM), which can manifest in either honest or deceptive forms (i.e., dishonest or faking) [2]. For example, job applicants may use honest IM to promote their accomplishments related to the job requirements, or they might use deceptive IM to fabricate answers to interview questions and claim nonexistent achievements to obtain a job offer [3].

While honest IM can improve selection validity, showcasing an applicant's self-presentation skills, deceptive IM risks injecting bias into hiring decisions [4]. Past research has found that even experienced interviewers cannot accurately detect and distinguish between interviewees' honest and deceptive IM behaviors (IMs), potentially distorting or inflating an interview evaluation [5]. This issue persists even in highly structured interviews [2]. The exploration of nonverbal valid deception signals and assessments in employment interviews is an area that requires further investigation, as called for by both psychologists [6] and computer scientists [7].

Psychological research suggests that facial expressions inherently serve as signals of emotional states. However, individuals do not always have full control over their facial muscles, which may inadvertently reveal their true emotional states. This phenomenon, known as emotional leakage, happens when the signals conveyed by facial expressions — intentionally or otherwise — unveil emotions that individuals are trying to hide or express more intensely or involuntarily than initially intended [8].

In certain situations, individuals may alter their expressions and head motions for image management and to conceal deceptive behaviors [9]. This phenomenon leads to the 'rigidity effect', wherein deceivers deliberately limit their facial expressions and head movements to appear truthful, particularly when under scrutiny or at risk of detection [10]. Consequently, the analysis of facial expressions and head movements becomes crucial in discerning both honest and deceptive behaviors.

A meta-analysis of 206 documents comprising 24,483 records [11] revealed that humans have a limited ability to distinguish between honest and deceptive expressions. The analysis revealed that humans correctly identified 61% of truthful messages as truthful and 47% of deceptive messages as deceptive.

With employment interviews, regardless of their experience, human interviewers can correctly recognize honest IMs at a rate of 13% to 29% and deceptive IMs at a rate of 12% to 19% [5].

This research was partially supported by the National Science and Technology Council (NSTC), Taiwan, under grants NSTC 109-2511-H-003-046, NSTC 110-2511-H-003-044-MY2, NSTC 112-2410-H-003-102-MY2, and NSTC 111-2410-H-019-006-MY3.

*Corresponding authors: Yu-Sheng Su (ccucssu@gmail.com)

Roulin and Powell [12] suggest that human interviewers struggle to accurately identify IMs, as they often rely on invalid cues from interviewees, such as anxiety. Dishonest interviewees tend to suppress their expressions more than honest ones to conceal their anxiety.

Research in psychology has revealed that even trained professionals struggle to accurately interpret the emotions of individuals based solely on legitimate cues from facial expressions. This challenge in discerning genuine and deceptive behaviors arises because reliable cues in facial expressions often occur in situations involving mixed or intricate deceptive strategies, which are not easily assessed through isolated cues. Furthermore, many of these reliable cues in facial expressions are too fleeting and subtle to be captured by the human eye. As a result, the limited attention span and memory capacity of human observers hinder their ability to identify and remember all the potential cues present in an individual's facial expressions [13].

Alternatively, an artificial intelligence (AI) facial expression detector, enabled by deep learning (DL), can augment the human ability to effectively discriminate between honest and deceptive IMs [14]. These methods are widely applied for automatically detecting deception from videos in high-stakes situations [7], [15]–[18], such as in the context of employment interviews. To achieve cost savings and avoid human interaction during the pandemic, asynchronous video interview platforms (AVIs) have been increasingly implemented by employers to screen job candidates [19]. AVIs powered by DL provide a resourceful channel to extract features for detecting job applicants' IMs [20], [21].

The use of facial expressions, analyzed by computer vision to detect deception, has been a widely debated topic. Previous studies have linked specific emotions to facial action units (AUs), which correspond to different facial muscles controlling aspects of movement. However, the reliability of AUs as indicators of emotions has been questioned due to potential influences from cultural norms, individual differences, and context [22].

Due to the lack of real-time interaction between job applicants and interviewers in AVIs, and considering that research on IMs is largely limited to face-to-face contexts, the manifestation of candidate IMs in AVIs, as well as how AI technology integration in AVIs influences interview evaluations, has been a topic calling for research [19], [23], [24].

To date, there has been no empirical research on the detection of IMs in AVIs by AI in real-life scenarios. Furthermore, no study has yet compared the effectiveness of AI with that of human judges in identifying IMs through facial expressions and head movements in high-stakes situations.

To address these limitations and build upon previous work, this research utilized computer vision and DL to identify and differentiate self-reported honest and deceptive IMs during high-stakes employment interviews from video sequences on an AVI platform. Unlike previous studies, this approach captured temporal patterns of facial expressions and head movements, rather than relying solely on AUs or human-coded emotion states. Furthermore, we assessed the validity of the AI detector versus professional human interviewers in identifying self-reported honest and deceptive IMs across multiple dimensions.

The remainder of this paper is organized as follows: Section II presents the related works for IM behavior detection using facial expressions from videos. Section III introduces the proposed methodology of our experiment for developing an AI detector that can identify various IMs in AVIs. Section IV demonstrates and analyzes the results of the experiment and the accuracy of IM identification by our AI detector and human interviewers. Section V discusses the findings and future work for these techniques and practices.

## II. Background and Related Works

### *A. IM as a Signaling Game in an Employment Interview*

Building on signaling theory [25], job interviewees attempt to convey positive and negative messages to present their suitability, while interviewers or raters interpret these signals to make hiring recommendations [26]. These signals can indicate either honest or deceptive IM [4]. Honest IM involves attractively presenting oneself through qualifications (honest self-promotion), praising the interviewer (honest ingratiation), and recounting actions taken to prevent negative occurrences (honest defensiveness). On the other hand, deceptive IM includes subtly crafting an image close to the truth of being a good candidate (slight image creation), extensively fabricating an image that includes false information (deceptive extensive image creation), defending an image by omitting negative information (deceptive image protection), and attempting to please the interviewer to gain favor, regardless of the selection criteria (deceptive ingratiation).

Although job interviewees may intentionally convey signals for both honest and deceptive IMs, signal receivers often attend to both the intended signals and unintended ones. These can include nonverbal deception cues such as the interviewees' rigidity and the temporal patterns of their facial expressions and head movements [27].

### *B. Facial Expressions and Head Movements as Signals for Identifying IM*

Facial expressions can be categorized into two distinct patterns: micro and macro expressions. Microexpressions last less than 1/2 of a second and usually last only 1/15 to 1/25 of a second [28], which reveals that a person is trying to mask or suppress some signals in his or her facial expressions [29]. Some deception studies have studied macroexpressions (lasting >=0.5 seconds) rather than microexpressions (lasting <0.5 seconds) and found fewer cues or signals to predict human deception from facial expressions [13], [30], [31]. In most cases, microexpressions are more informative and ecologically valid for detecting deception than macroexpressions. [32].

According to Warren et al. [13], "subtle expressions" can serve as reliable indicators for identifying deception and are comparable or even superior to microexpressions. Subtle expressions occur infrequently in mini regions of the face when a person is lying and are not easily detected by human eyes because of the restricted and small regions and their infrequent manner (i.e., "subtlety"), although not because of the duration of the facial muscle movement [33]. Thus, subtle expressions may contain both macro- and microexpressions [34].

Although subtle and microexpressions are reliable and valid signals for discriminating between honest and dishonest individuals in labs using supported instruments [32], [35], most human observers have cognitive loads or coding limitations that prevent them from reading these signals because subtle and microexpressions are indistinct, temporal, and short-lived [36].

Computers and humans have vastly different abilities to detect deception [37]. To improve human detection ability, Ekman *et al*. [38] developed the facial action coding system (FACS), which utilizes all possible visible signals of facial muscle movements, including subtle and microexpressions. This approach depends on human experts to label the AUs related to universal emotions from static images, which can be biased and constrained by a universal framework that overlooks cultural variations [39]. Additionally, the use of static images may not fully capture the complexity and inconsistency of individual subtle- and microexpressions in specific stimuli and contexts, which can hamper detection reliability in related works [22], [40].

Studies in cognitive science show that deception signals are rich but subtle and discrete across multiple timescales, with varying degrees of pattern and regularity. Instead, the temporal patterns of facial expressions should be fully captured to identify more valid and discrete signals that can distinguish honest and deceptive IMs [22], [40]. Given that to achieve the goal of capturing temporal patterns of facial expressions, including macro-, subtle-, and microexpressions, as well as head movements, a more effective methodology would be to use sequence images instead of static images from videos and employ computer vision-based motion profiles. By leveraging computer vision and neural networks (NNs), it is possible to analyze and decode the complex, temporal patterns of facial expressions and head movements.

### *C. AI Modeling to Decode the Signals*

Wu et al. [18] suggested using computer vision to capture and decode the temporal patterns of facial expressions, aiming to create an automatic deception detector for video applications. In addition, Twyman *et al.* [41] proposed that computer vision enabled with DL via neural networks (NNs) can achieve better detection accuracy than human observers in field online automated interview settings, such as in AVIs.

The iBorderCtrl Intelligent Portable Control System is the first known attempt to utilize computer vision and NNs to develop an AI model that can evaluate the deception risk of third-country national travelers crossing EU land borders. The system works by analyzing passengers' nonverbal signals, including facial expressions and head movements, during an AI interview conducted by a virtual agent, Avatar. The system was developed based on the "silent talker" concept, which involves asking passengers a series of security questions (e.g., "What is your citizenship and the purpose of your trip?") and assigning a truthfulness score out of 100. Funding for this project was provided by the European Union's (EU) Horizon research and innovation program [42].

The original silent talker approach had an accuracy rate of only 43-54% in identifying deception, which was no better than human judgment [11]. The developers of iBorderCtrl fine-tuned the system and subsequently trained and tested a sample of 32 volunteer participants. The results demonstrated that the system could achieve an overall accuracy rate of 75.55% for identifying honest interviewees and 73.66% for detecting deceptive ones [43].

Hereafter, several studies have subsequently introduced AI deception works based on facial expressions using NNs and videos from real-life court trials [7], [15]–[18]. Although facial expressions are more effective cues for detecting deception than body language or speech analysis [44], the labeling of facial expressions in these studies relied on human hand-crafted features. These include AUs or human-coded emotion states associated with inferred deception, rather than deception itself [45]. In employment interviews, the range of IMs is varied, and they cannot be solely categorized as honest or deceptive [4].

To overcome the shortcomings identified in previous studies, our study employed computer vision and NNs to capture the temporal patterns of facial expressions and head movements as deep features [46]. These features enable the AI discrimination of self-reported honest and deceptive IMs during real employment interviews, using image sequences from videos on an AVI platform. Additionally, we compared the efficacy of our AI detector with that of professional human interviewers in accurately identifying self-reported honest and deceptive IMs across various dimensions.

## III. METHODS

### *A. Data Collection and Data Labeling with IM*

The study adopted a field experimental design in which we contracted a Taiwanese company that uses an AVI platform based on an agreement of industry-academia cooperation. The company and its subsidiaries screen their job applicants by the AVI in the initial stage of pre-employment assessment. The company's HR department helped us by sending invitation emails to their job applicants who applied to fill a job vacancy in the company on the AVI. The participants responded to us through an online survey system without the company's knowledge. We solicited 150 participants to build a real-life dataset based on our research funding. Twenty-nine job applicants' data were eliminated from this study because they did not complete the recorded video interview or questionnaires thoroughly. The remaining 121 job applicants were from various industrial sectors (Public service 2.48%, Administration 2.48%, Communication 4.13%, Hospitality 0.83%, Retail marketing 16, Customer service 12, Finance 14, Technical service 13.22%, Education & training 9.09%, Logistic 2.48%, Manufacturing 8.26%, Construction 1.65%, Medical health 3.31%, Entertainment 4.13%, and Others 13.22%); genders (Male 46.28%, Female 53.72%), educational degrees (Doctorate 0.83%, Master 26.44%, Bachelor 72.73%), and positions (Manager 21.49%, non-Manager 78.51%). Their average age and average years of work experience were 30.37 and 5.41, respectively.

Everyone who applied to the company was invited to log into the AVI software page, after which the interview process began automatically. Each question was displayed on a single screen of the interviewees' mobile devices, with a three-minute time limit to respond before the software automatically progressed to the next question page. If the interviewee finished the question within three minutes, he/she could press the next question button.

Interviewees faced five structured behavioral questions crucial for different job roles, focusing on assessing interpersonal communication skills. These questions drew on past experiences and avoided hypothetical or future-oriented scenarios (e.g., "Can you share an instance when you had to simplify and explain a complex procedure or task to an individual or a team?"). Using behavioral-based structured interview questions was expected to trigger the interviewees' honest self-promotion, honest defensiveness, deceptive image creation (lying), and deceptive image protection [4]. The entire interview process for each participant took approximately 15 minutes to complete. The hiring process in the AVI protocol did not permit requests for reviewing and rerecording.

Following the interview, participants were sent an email invitation to participate in the study. Participants who agreed to participate completed an online questionnaire on self-reported honest and deceptive IMs and signed an electronic consent form. The authors ensured that all personal information would be kept confidential and protected and that the data collected would be used only for academic research purposes. They also guaranteed that the information would not be disclosed to any parties that could potentially impact employers' hiring decisions. This approach was preferred because self-reported measures are considered more reliable than human observations when the reporter is not motivated to fake their answers [47]

Bourdage et al.'s [4] short five-point scale was used to assess honest and deceptive self-reported IMs. This scale comprises 4 items each for honest self-promotion, honest defensiveness, deceptive extensive image creation, and deceptive image protection, along with an additional 2 items for deceptive slight image creation, totaling 18 items. The original questions about honest ingratiation, deceptive ingratiation, and two specific questions about slight image creation ("I distorted my answers based on the comments or reactions of the interviewer" and "I distorted my answers to emphasize what the interviewer was looking for") [4] [20] were excluded. These questions were omitted as they require two-way interviewee-interviewer interactions, which are not applicable in one-way AVIs.

To validate our DL modeling and compare it with the ability of human interviewers to identify honest and deceptive IM, we recruited 30 HR professionals from a Taiwanese HR association, each with over three years of experience in employment interviewing. These interviewers assessed the IMs of 30 interviewees using an additional testing set of 30 video data sources, which were separate from the 121 sources used for modeling. A total of 24 female (80%) interviewers (average age: 35.8 years; average years of permanent job experience: 9.9 years) and 6 male (20%) interviewers (average age: 34.2 years; average years of permanent job experience: 8.3 years) participated in the study.

The 30 human interviewers were asked to rate the honest and deceptive IMs of three interviewees each based on the 18-item questionnaire using the 30 video data sources. Likewise, each interviewee was evaluated by three human interviewers based on their IMs, as observed by the interviewers.

The 30 videos were recorded by real job applicants from various industrial sectors collaborating with this study at the same company. These sectors included Public Service (3.33%), Administration (3.33%), Communication (3.33%), Hospitality (3.33%), Retail Marketing (2), Customer Service (6.67%), Finance (2), Technical Service (6.67%), Education & Training (3.33%), Logistics (3.33%), Manufacturing (3.33%), Construction (3.33%), Medical Health (3.33%), Entertainment (3.33%), and Others (6.67%). The participants comprised 14 males (46.67%) and 16 females (53.33%) with educational qualifications of Master's degrees (5, 16.67%) and Bachelor's degrees (25, 83.33%), and professional positions of Manager (5, 16.67%) and Non-Manager (25, 83.33%). The average age of the participants was 30.95 years, with an average work experience of 6.25 years.

### *B. Analyzing Temporal Facial and Head Movement Patterns*

We obtained the initial 121 interview videos and transformed them into image sequences by employing OpenCV. This approach enabled us to capture the macro-, subtle-, and microexpressions, as well as head movements of each participant frame-by-frame from the videos, which were recorded at 30 frames per second (fps), as employed by Qu *et al.* [48]. To minimize the variability of the image frames resulting from rotation and shifting, we resized the frame images to a uniform width of 640 pixels [49]. The video recordings from the AVI software used in this study were captured at a resolution of 720 pixels, and the majority of the job applicants' smartphones captured video at a resolution of 720 pixels and a maximum of 30 fps [50].

To capture the temporal patterns of the interviewees' facial expressions and head movements, which included facial movements, head nods, head shakes, and head twists, we identified the facial landmarks in each image sequentially at 30 fps [51]. This process involved detecting the regions of interest (ROIs) on the face and the corresponding facial deformations, rather than merely counting static AUs. We utilized the Google MediaPipe FaceMesh library, capable of estimating 468 3D facial landmarks in real-time on mobile devices without the need for dedicated depth sensors. This tool is effective for identifying facial positions to estimate transformations within a metric 3D space [52], as shown in Fig. 1.

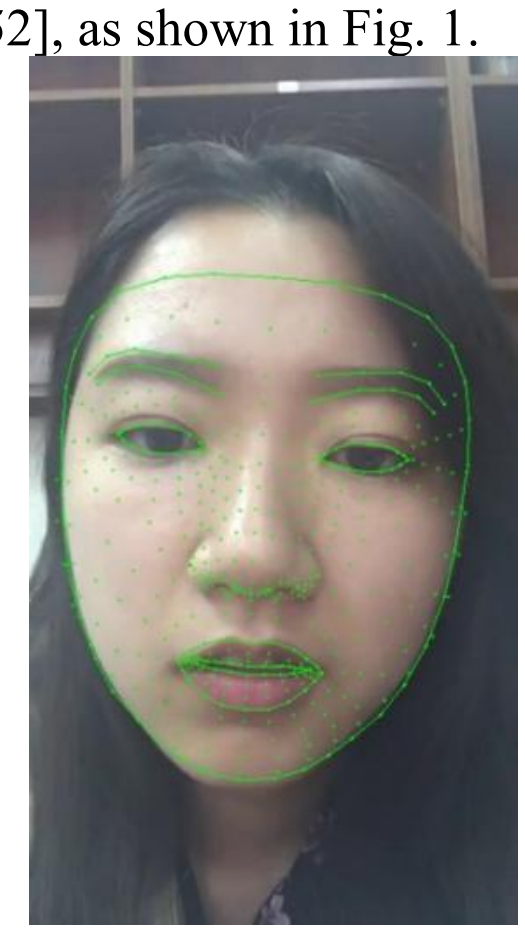

**Fig. 1.** Face mesh topology with facial landmarks.

The MediaPipe Face Mesh utilizes two interconnected NN models to deduce the 3D surface geometry of a human face from a single smartphone or webcam input, eliminating the need for specialized depth sensors. One model processes the entire image to identify face locations, while the other focuses on these located faces to predict their 3D surface geometry using regression [53].

To ensure precise identification of facial landmarks even when interviewees move their heads, preprocessing techniques such as localization, translation, and rotation are employed to crop and detect face images. Subsequently, the color images are converted to grayscale to reduce noise and normalize the large-scale images.

As illustrated in Fig. 2, we established a horizontal axis (in orange) to align the corners of the eyes and then utilized the rectangle of 3D surface geometry to detect angle and proportion changes along this axis.

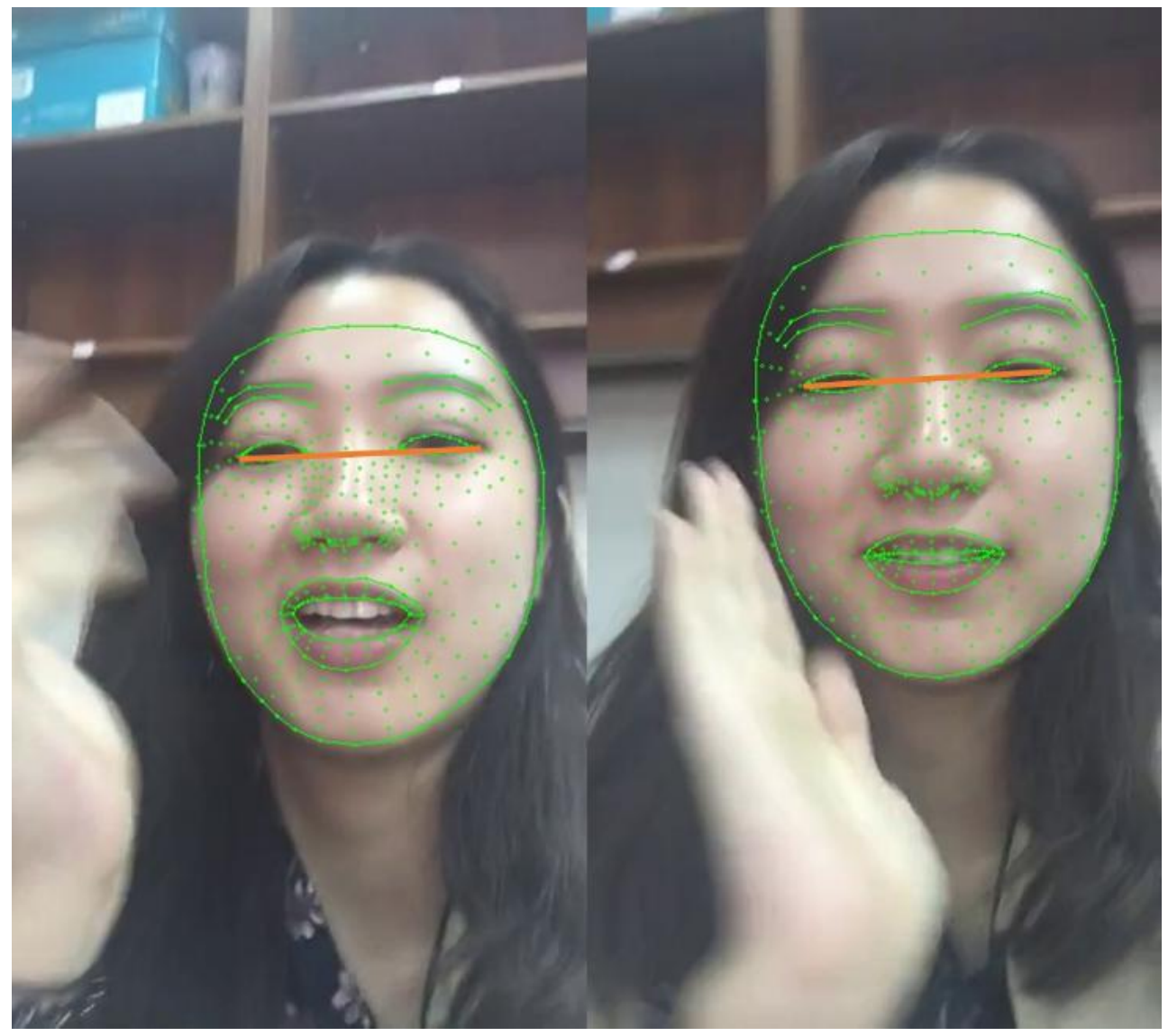

**Fig. 2.** Eye corner alignment

To convert the original 2D images into a vector of 3D coordinates, we transformed these images into 3D space using a Cartesian coordinate system [54], as shown in Fig. 3.

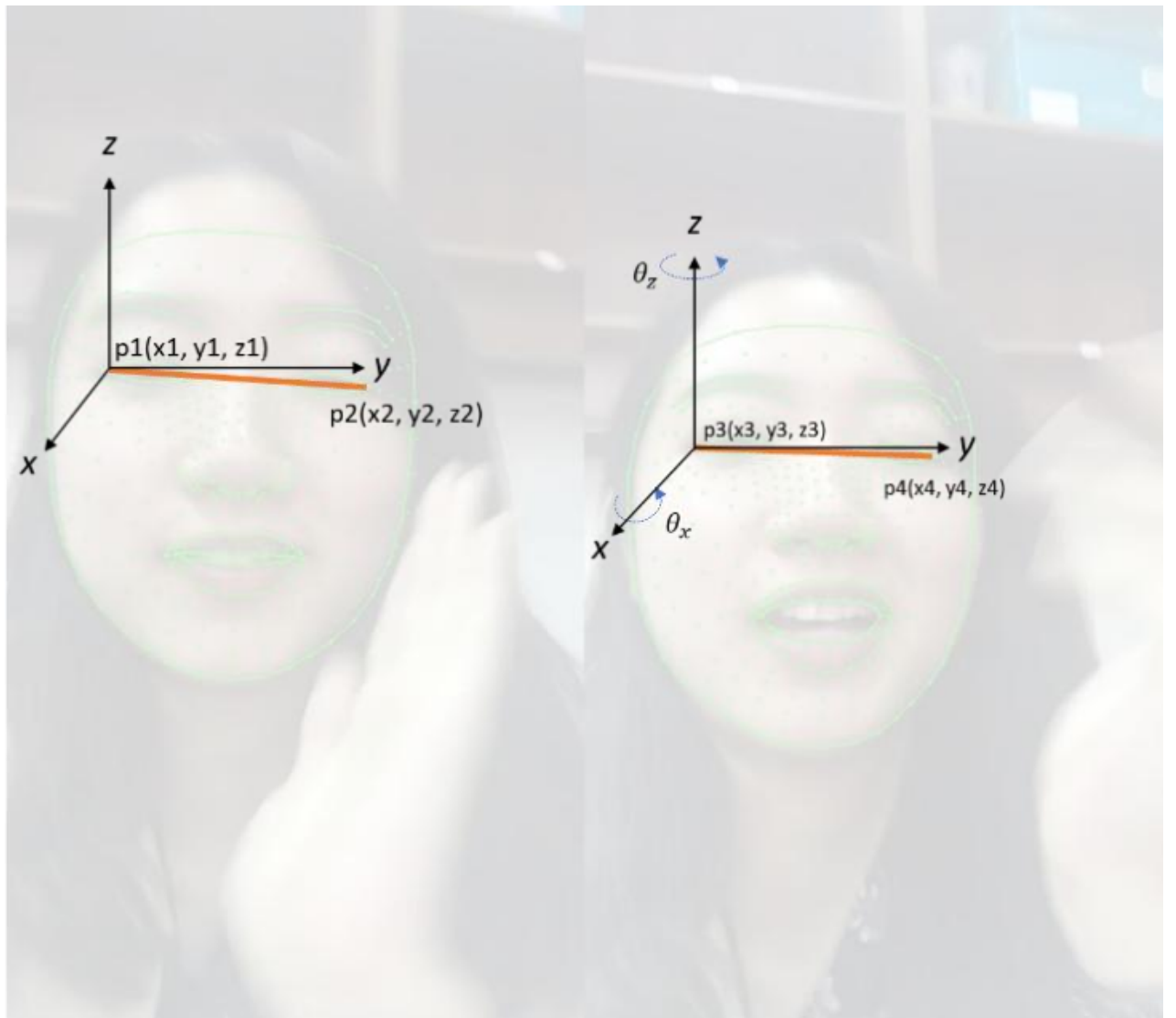

**Fig. 3.** 2D to 3D image transformation

We obtained $P(X_{fm}, Y_{fm}, Z_{fm})$ from FaceMesh, where the y-coordinate matches the previously mentioned horizontal axis. The x and y coordinates of the vertices match the points in the 2D image. We rotated the 2D images along the x-, y-, and z-axes using the Cartesian coordinate system. The height ($h$) and width ($w$) of the image were calculated as depicted in equation (1), while the 3D Cartesian space was defined by a transformation matrix (2).

$$f = \sqrt{h^2 + w^2} \tag{1}$$

$$\begin{bmatrix} x \\ y \\ z \end{bmatrix} = \begin{bmatrix} 1 & 0 & -w/2 \\ 0 & 1 & -h/2 \\ 0 & 0 & 1 \\ 0 & 0 & 1 \end{bmatrix} \begin{bmatrix} 1 & 0 & 0 & 0 \\ 0 & 1 & 0 & 0 \\ 0 & 0 & 1 & 0 \\ 0 & 0 & 0 & 1 \end{bmatrix} \begin{bmatrix} f & 0 & w/2 & 0 \\ 0 & f & h/2 & 0 \\ 0 & 0 & 1 & 0 \end{bmatrix} \begin{bmatrix} X_{fm} \\ Y_{fm} \\ Z_{fm} \end{bmatrix} \tag{2}$$

After transforming the 3D coordinates, we processed and computed the images as follows: each image's timing was denoted as (t), and the image at a specific time was referred to as ($frame[t]$). The corner of the left eye in the image at a specific time was set as $p1(x1, y1, z1)$, and the corner of the right eye at the same time was set as $p2(x2, y2, z2)$, forming a line *L1* between *p1* and *p2*. In the subsequent time series, $frame[t+1]$, we formed line L2 connecting points $p3(x3, y3, z3)$ and $p4(x4, y4, z4)$. The image sequences in $frame[t+1]$ were then rotated along the *x*-, *y*-, and *z*-axes. The angles along these axes were computed as equations (3), (4), and (5), respectively. $\theta_{xy}$ represents a head twist along the z-axis and can be computed using matrix $R_z$. $\theta_{xz}$ represents a head nod along the y-axis and can be computed using matrix $R_y$. $\theta_{zy}$ represents a head shake along the x-axis and can be computed using matrix $R_x$.

$$\theta_{xy} = \tan^{-1}\left(\frac{|y_2 - y_1|}{|x_2 - x_1|}\right) \cdot \frac{180}{\pi} - \tan^{-1}\left(\frac{|y_3 - y_4|}{|x_3 - x_4|}\right) \cdot \frac{180}{\pi} \tag{3}$$

$$\theta_{xz} = \tan^{-1}\left(\frac{|z_2 - z_1|}{|x_2 - x_1|}\right) \cdot \frac{180}{\pi} - \tan^{-1}\left(\frac{|z_3 - z_4|}{|x_3 - x_4|}\right) \cdot \frac{180}{\pi} \tag{4}$$

$$\theta_{zy} = \tan^{-1}\left(\frac{|y_2 - y_1|}{|z_2 - z_1|}\right) \cdot \frac{180}{\pi} - \tan^{-1}\left(\frac{|y_3 - y_4|}{|z_3 - z_4|}\right) \cdot \frac{180}{\pi} \tag{5}$$

Subsequently, we derived $\theta_{xy}$, $\theta_{xz}$ and $\theta_{zy}$. The rotation matrices $R_x$, $R_y$ and $R_z$ were defined by rotation matrices (6), (7), and (8), respectively. Matrix (6) was used to describe the scenario of an object rotating around the *x*-axis, such as the head nodding up and down. Matrix (6) was used to describe the scenario of an object rotating around the x-axis, like the head shaking left and right. Matrix (7) captured the rotation around the y-axis, such as the head nodding up and down. Matrix (8) depicted rotation around the z-axis, such as looking left or right or twisting the head. In these matrices, $cos(\theta)$, and $sin(\theta)$ represented the cosine and sine values of angle $\theta$, respectively. These matrices allowed us to precisely apply rotation operations to points or objects in three-dimensional space, altering their orientation without changing their positions.

$$R_x = \begin{bmatrix} 1 & 0 & 0 & 0 \\ 0 & cos(\theta_{zy}) & -sin(\theta_{zy}) & 0 \\ 0 & sin(\theta_{zy}) & cos(\theta_{zy}) & 0 \\ 0 & 0 & 0 & 1 \end{bmatrix} \tag{6}$$

$$R_y = \begin{bmatrix} cos(\theta_{xz}) & 0 & -sin(\theta_{xz}) & 0 \\ 0 & 1 & 0 & 0 \\ sin(\theta_{xz}) & 0 & cos(\theta_{xz}) & 0 \\ 0 & 0 & 0 & 1 \end{bmatrix} \tag{7}$$

$$R_z = \begin{bmatrix} cos(\theta_{xy}) & -sin(\theta_{xy}) & 0 & 0 \\ sin(\theta_{xy}) & cos(\theta_{xy}) & 0 & 0 \\ 0 & 0 & 1 & 0 \\ 0 & 0 & 0 & 1 \end{bmatrix} \tag{8}$$

To maintain a consistent distance between the eyes after rotation, we used the distance from $p1(x1, y1, z1)$, to $p2(x2, y2, z2)$ as a basis, and the distance from the rotated $p3(x3, y3, z3)$ to $p4(x4, y4, z4)$, to calculate the scaling ratios for $x$, $y$, $z$ as shown in matrix (9).

$$S_c = \begin{bmatrix} \frac{|x2-x1|}{|x4-x3|} & 0 & 0 & 0 \\ 0 & \frac{|y2-y1|}{|y4-y3|} & 0 & 0 \\ 0 & 0 & \frac{|z2-z1|}{|z4-z3|} & 0 \\ 0 & 0 & 0 & 1 \end{bmatrix} \quad (9)$$

After scaling, the image might not fit the original canvas. Thus, we adjusted the $x$, $y$, and $z$ axes using shearing matrices (10), (11), and (12). Since the rotation was counterclockwise, we converted the angle $\theta$ to $\varphi$, with $\varphi = \theta - 360°$.

$$Sh_x = \begin{bmatrix} 1 & 0 & 0 & 0 \\ tan(\varphi_y) & 1 & 0 & 0 \\ tan(\varphi_z) & 0 & 1 & 0 \\ 0 & 0 & 0 & 1 \end{bmatrix} \quad (10)$$

$$Sh_y = \begin{bmatrix} 1 & tan(\varphi_x) & 0 & 0 \\ 0 & 1 & 0 & 0 \\ 0 & tan(\varphi_z) & 1 & 0 \\ 0 & 0 & 0 & 1 \end{bmatrix} \quad (11)$$

$$Sh_z = \begin{bmatrix} 1 & 0 & tan(\varphi_x) & 0 \\ 0 & 1 & tan(\varphi_y) & 0 \\ 0 & 0 & 1 & 0 \\ 0 & 0 & 0 & 1 \end{bmatrix} \quad (12)$$

### *C. Modeling IM Based on 3D-CNN and LSTM*

In our proposed method, we used a three-dimensional convolutional neural network (3D-CNN) [55] as the DL algorithm to estimate two honest IMs and two deceptive IMs because CNN is a widely used, high-performing, and uncomplicated method for recognizing facial and head movement features [56]. The 3D-CNN combines a CNN with a temporal dimension to process spatiotemporal data from video sequences. On average, each candidate's total video recording was approximately 15 minutes long. We split them into 5-minute segments and recorded a new segment every minute, resulting in a 4-minute overlap with the previous segment according to the "motion estimation" approach [57]. Each participant provided 11 video clips, resulting in a total of 1,331 short video clips used for each model.

Our modeling was built on Python (version 3.7.13) using the TensorFlow framework (version 2.10.0) with the Keras library (version 2.10.0). We utilized a 3D-CNN model to process the videos and used long short-term memory (LSTM) to extract features with time sequences. These features were then fed into the regression head, which outputted values in the range of [1, 5] as the IM scale, as shown in Fig. 4. The dataset underwent random division into training, validation, and testing sets with an 80-10-10 ratio [58]. The process aimed to train, select, and generalize the four models that can automatically estimate the four IM scores using the extracted temporal features of facial expressions and head movements.

The models' inputs were the entire interview video sequence frames, including fps×image pixels×3 RGB color channels, and processed facial landmarks to emphasize the ROIs for the 3D-CNN. The output was the IM scores across four IM dimensions.

The network consisted of several fixed convolutional 3D layers (Conv3D) and pooling layers (Pool). Conv3D summarizes the features in an input image by applying learned filters. A pooling layer was then added to downsize the features in patches of feature maps created by Conv3D. To address the vanishing gradient issue linked to sigmoid and hyperbolic tangent activation functions, Conv3D applied a rectified nonlinear unit to the feature maps [59].

As depicted in Fig. 4, the 640×640 input images with 30 fps and 3 color channels (30×640×640×3) were initially filtered by Conv3D with random weights. The images were then reduced to 30×320×320×32 by the Conv3D_1 layer. Subsequently, the Pool_1 layer downsized the image to 30×160×160×64. This process was repeated from the Conv3D_2 layer to the Pool_2 layer and from the Conv3D_3 layer to the Pool_3 layer, resulting in feature maps and an image size of 30×10×10×256. The Padding layer expanded the image size to 30×30×30×256 from the Pool_3 layer. The Conv3D_4 layer further reduced the image size to 30×8×8×512. The Pool_4 layer then minimized the feature tensor obtained in Conv3D_5 to an image size of 30×2×2×2048. The final convolution reduced the image to 30×1×1×4096 with a 1×1 filter through average pooling to compute each patch on the feature map. The fully connected layer computed the IM scores and delivered them to the Long Short-Term Memory (LSTM) layer [60] with 512 units. Eventually, the output of the LSTM layer was sent to the regression head to estimate each IM score.

The regression head included 4 dense layers and performed 3D-CNN network computations from input features to output data, applying the LSTM layer followed by ReLU activations. The input to the Dense_1 layer began with 512 units from LSTM. The process from the Dense layer to Dropout was repeated until Dense_4 was fully connected with 128 neurons. A dropout layer was included between each pair of dense layers at a rate of 0.5 to prevent model overfitting [61]. During the training process, we iterated a thousand times, with each batch consisting of 4 samples, a learning rate of 0.001, and an evaluation frequency of every 10 batches.

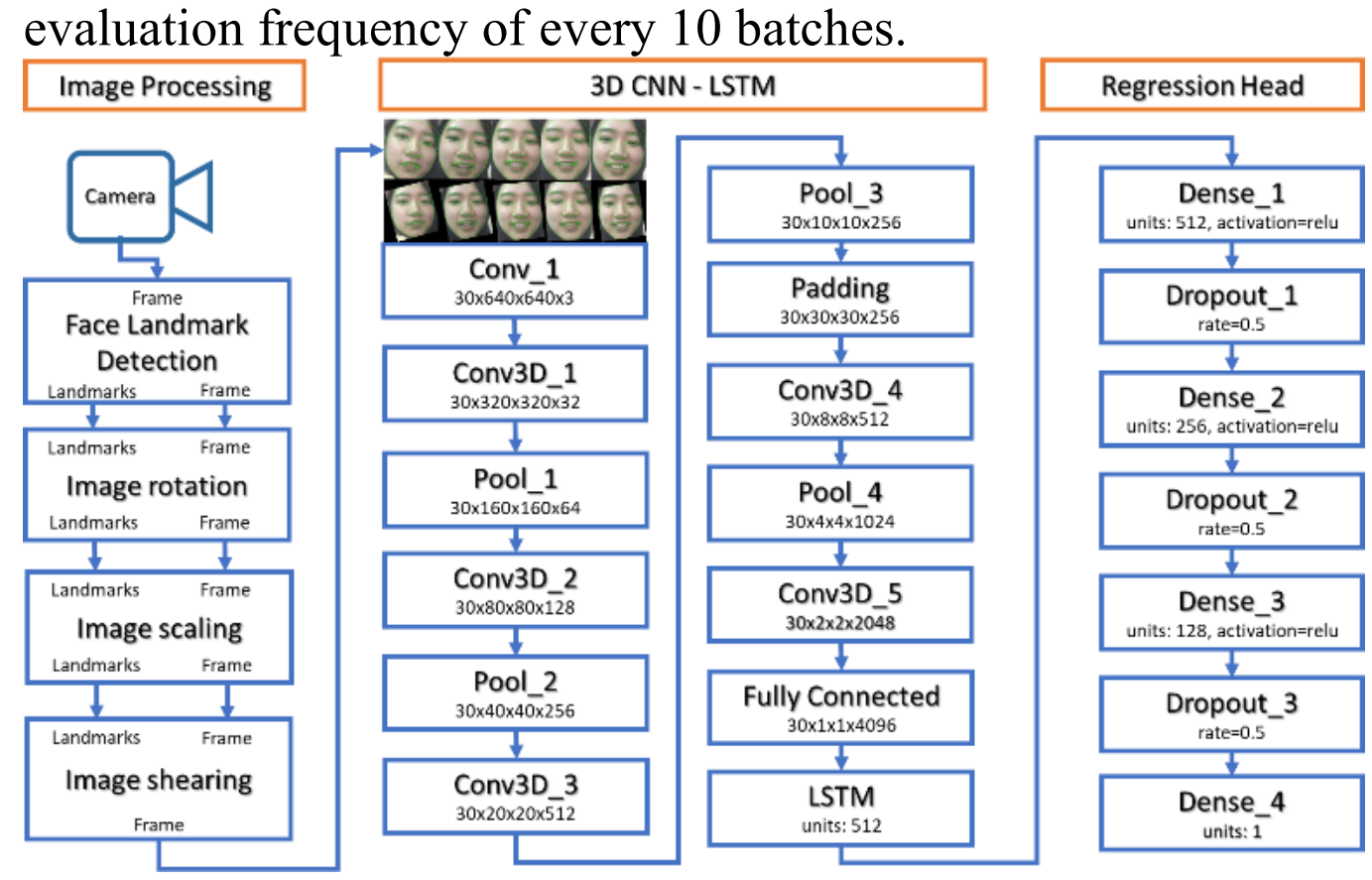

**Fig. 4.** Proposed Image Processing and 3D-CNN - LSTM Regression Models

## IV. RESULTS

### *A. Self-reported IM: Statistics, Validity, and Reliability*

To assess the construct validity and reliability of our model for self-reported IMs — including honest self-promotion, honest defensiveness, deceptive slight/extensive image creation, and deceptive image protection — which consist of 18 items from Bourdage et al.'s [2] IM scales, we translated these scales from English into Mandarin Chinese. We then conducted an exploratory factor analysis (EFA) with varimax rotation (eigenvalues > 1) and performed an internal consistency analysis using Cronbach's alpha.

The EFA indicated that the two items of deceptive slight image creation and the four items related to deceptive extensive image creation converged under a single dimension. As a result, this dimension has been designated as 'Deceptive image creation'. The remaining three dimensions were found to be in alignment with those of the original scale.

Prior to performing the EFA, Kaiser-Meyer-Olkin (KMO) and Bartlett's test of sphericity were conducted to test whether our sampling was adequate for factor analysis. The results showed that the KMO value (0.84) was more than 0.5 with a significant Bartlett's test (<0.01), which indicated that factor analysis was appropriate for our sampling data [62].

The mean, standard deviation (SD), range (from minimum to maximum values), and mean score distributions for the IM dimensions are shown in Table I. The table reveals that, on average, applicants in this study demonstrated more honest behaviors than deceptive IMs in AVIs. Most applicants rated their honest IMs between 3.00 to 3.99, indicating 'agree', while the majority assessed their deceptive IMs in the 2.00 to 2.99 range, suggesting 'disagree'. Only a tiny proportion selected 'strongly disagree' for either honest or deceptive IMs. These findings imply that during the actual video-recorded interviews, applicants tended to exhibit some level of IMs.

Table I shows that all IM dimension factor loadings exceeded 0.5, meeting the convergence criteria with significant eigenvalues (>1) [62] and explaining 77.12% of the cumulative variance for IM across four distinct factors. Cronbach's alpha values for all dimensions were above 0.7, indicating adequate internal consistency for the four IM variables [63]. We then calculated the mean scores for questionnaire items within these dimensions to derive four IM factor scores.

Furthermore, we performed an analysis of variance (ANOVA) and Pearson correlation analysis to determine if the demographic characteristics of job applicants, such as industrial sectors, gender, educational degree, and positions, were associated with their self-reported IMs. No significant correlations were found between any demographic variable and specific IM scores. Levene's Test revealed no significant variance differences ($p > 0.05$) in the ANOVA, confirming the homogeneity of variances. Consequently, these demographic factors were not used as statistical controls in this study.

To assess the reliability of the self-reported IM measures, we employed an automatic personality recognition based on temporal patterns of facial expression modeling [64] to evaluate the Big Five personality traits of job applicants. We then conducted multiple linear regression (MLR) analyses four times to explore the relationship between participants' Big Five personality traits and their self-reported IMs across the four IM dimensions. The MLR models revealed that interviewees with higher recognized scores for neuroticism and lower scores for extraversion reported more deceptive image creation (standardized $\beta$ = .250 and -.247, respectively, $p < .05$), accounting for 13.8% of the explanatory power ($F = 3.835$, $p < .01$). In line with Bourdage et al.'s [4] empirical study, individuals low in extraversion tend to avoid using deceptive IM in job interviews due to concern about others' perceptions. Conversely, those high in emotional stability (the opposite of neuroticism) are more likely to conform to social desirability and engage less in deceptive IMs [65]. Our findings align with previous psychological research indicating that neuroticism is positively associated with lying, while extraversion is negatively associated with lying [66]. Based on this evidence, we assert that the self-reported IM measures are reliable.

### *B. Performance of the AI Models in IM Detection*

Our proposed models were validated using 121 video interviews, employing Pearson correlation coefficient (R), explained variation ($R^2$), and root mean square error (RMSE) as metrics. To examine the AI detector's concurrent validity, we collected a sample of another 30 video interviews from other participants in similar scenarios and conducted a Pearson correlation analysis between computer/human and self-reported IM measures to compare the coefficients. The additional data were kept separate and not included in the modeling.

Table II shows that we created distinct models for each of the four IM types, which achieved high correlation coefficients between 0.89 and 0.96. This achievement demonstrates a strong alignment between the IMs inferred by our models and the corresponding self-reported IMs, suggesting a high degree of concurrent validity. All four models showed correlation coefficients greater than 0.7, thus confirming the robust validity of our proposed modeling approach [62].

Additionally, our models exhibited remarkable explanatory power ($R^2$). They accounted for 91% and 84% of the variance in honest and deceptive IMs, respectively, and explained a significant proportion (79-92%) of the variance in self-reported IM scores during personnel selection in a real-life AVI. These results indicate that dynamic facial expressions and head movements are effective indicators for assessing the extent of various behaviors exhibited. Furthermore, our models significantly surpassed human professionals in evaluating interview videos, with the latter only achieving an $R^2$ of 13%-29% for honest IMs and 12%-19% for deceptive IMs [5].

To measure the error of predictive models, we used RMSE, where lower values indicate better performance. In our study, RMSE values ranged from 0.25 to 0.41. Since SD occurs naturally in null deviance models, it can be used as a benchmark for models to outperform. Our proposed framework achieved an average RMSE/SD of 0.36 for the four predicted IM metrics, indicating that RMSE reduced randomness better than SD.

### *C. Comparison of AI and Human Interviewers in IM Detection*

Fig. 5 presents four scatter plots illustrating the relationship between estimated scores and job candidates' self-reported scores for the four IMs. These data were sourced from 30 interview videos not included in our modeling process and were evaluated by both human interviewers and our AI detection models. To assess the concurrent validity of the job candidates' self-reported IMs, we utilized the correlation coefficient (R) for comparing the results from human interviewers with those from AI modeling.

TABLE I
DESCRIPTIVE STATISTICS, CONSTRUCT VALIDITY, AND RELIABILITY FOR IM BEHAVIORS

| IM behaviors | Means | SD | Range | Means Distributions | Factor loading | Eigenvalues | % of variance | Cronbach's alpha |
|---|---|---|---|---|---|---|---|---|
| Honest self-promotion | 3.54 | 0.87 | 1-5 | 1.00-1.99 (4, 3.31%)<br>2.00-2.99 (17, 14.05%)<br>3.00-3.99 (55, 45.45%)<br>4.00-5.00 (45, 37.19%) | 0.82-0.91 | 3.21 | 20.06 | 0.92 |
| Honest defensiveness | 3.37 | 0.86 | 1-5 | 1.00-1.99 (4, 3.31%)<br>2.00-2.99 (25, 20.66%)<br>3.00-3.99 (59, 48.76%)<br>4.00-5.00 (33, 27.27%) | 0.74-0.87 | 2.88 | 17.98 | 0.86 |
| Deceptive image creation | 2.83 | 0.90 | 1-5 | 1.00-1.99 (5, 4.13%)<br>2.00-2.99 (68, 56.20%)<br>3.00-3.99 (26, 21.49%)<br>4.00-5.00 (22, 18.18%) | 0.76-0.91 | 3.13 | 19.54 | 0.90 |
| Deceptive image protection | 3.03 | 0.86 | 1-5 | 1.00-1.99 (6, 4.96%)<br>2.00-2.99 (50, 41.32%)<br>3.00-3.99 (41, 33.88%)<br>4.00-5.00 (24, 19.83%) | 0.80-0.85 | 3.13 | 19.53 | 0.89 |

TABLE II
MODELING RESULTS

| IM behaviors | R | $R^2$ | RMSE |
|---|---|---|---|
| Honest self-promotion | 0.96 | 91.79% | 0.25 |
| Honest defensiveness | 0.95 | 89.57% | 0.28 |
| Deceptive image creation | 0.94 | 88.62% | 0.31 |
| Deceptive image protection | 0.89 | 78.82% | 0.41 |

Before conducting the comparative analysis, we performed ANOVA (Levene's Test $p > 0.05$) and correlation analysis to assess the impact of human interviewers' gender, age, and work experience on their ratings of the four IMs. However, we found no statistically significant effects, leading us to exclude their demographic data as control variables. Additionally, we observed no significant differences in the four self-reported IM scores between the 30-video experimental dataset and the 121-video modeling dataset. Consequently, we can apply our modeling framework to another sample and retest the validity of the four models.

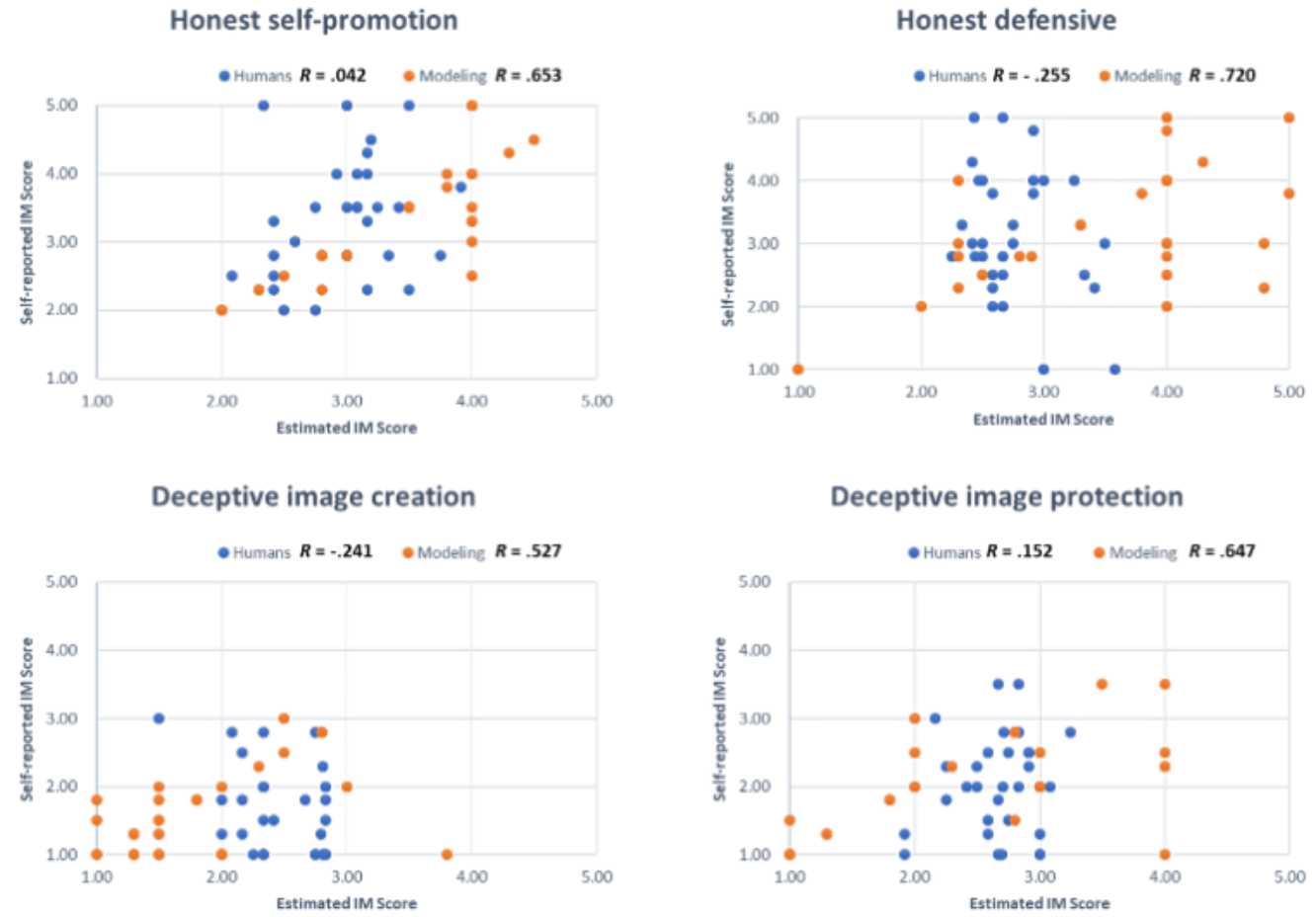

**Fig. 5.** Estimated vs. Self-Reported IMs: Scatter plots and coefficients.

Our model, using 30 samples, showed a stronger correlation (R = 0.53 to 0.72) between AI-estimated and self-reported IM scores of job applicants than human interviewers did (R = -0.26 to 0.15). In certain dimensions, such as honest defensiveness and deceptive image creation, the judgments of human interviewers contradicted the subjects' self-reported measures.

Additionally, we carried out an intraclass correlation coefficient (ICC) analysis using a two-way mixed-effects model for average measures of consistency. This analysis revealed that the interrater reliabilities of human interviewers for the four IMs were all below 0.2, indicating a lack of consistency and reliability among human interviewers in assessing IMs [67].

These findings are in line with results from previous psychological experiments, which show minimal significant correlation in field studies between job applicants' self-reported IMs and interviewers' perceptions of these IMs [36]. Additionally, they support the notion that experienced human interviewers struggle to detect either truth or deception during employment interviews [5], [12]. Furthermore, the findings raise concerns about the labels of truth or false given by human observers as a ground truth when modeling deception detectors.

As attested by Twyman *et al.* [41] and Wu *et al.* [18], our proposed AI detector applying Motion Estimation and 3D-CNN plus LSTM can augment the human capability to identify honest self-promotion, honest defensiveness, deceptive image creation, and deceptive image protection in real-life employment AVIs. This is achieved by analyzing the temporal features of interviewees' facial expressions and head movements using DL, going beyond the limitations of human annotation of expression signals.

## V. DISCUSSION

The paper's key contributions and limitations are outlined below. Firstly, we introduce a straightforward image-processing methodology that leverages existing open-source tools (Google's FaceMesh, TensorFlow, and Keras), eliminating the need for additional annotations or programming. This methodology detects temporal patterns of facial expressions and head movements by segmenting video clips into a series of single frames with facial landmarks. This segmentation, using 'motion estimation,' enables AI to detect dynamic signals of emotional leakage and the rigidity effect in job candidates' IMs, setting it apart from traditional static methods such as AUs or human-coded emotion states. For future research, this methodology can be replicated and adapted to develop a deception detector at the discretion of researchers.

Second, our study was conducted in a real-life personnel selection context using an AVI platform, as opposed to a laboratory setting. To our knowledge, this is the first attempt to use 3D-CNN and LSTM to detect actual job candidates' IMs by analyzing the temporal patterns of their facial expressions and head movements. However, the focus of this study on Chinese individuals in a competitive job market may not directly apply to different cultures or environments with less competition, where deceptive IM is less common compared to that in the U.S. or Chinese communities [36], [68].

Third, in contrast to absolute true and false labeling of deception in the previous computer science studies, this study adopted Likert scales to measure the extent of a job candidate's IMs based on personnel psychology rather than simply measuring whether job candidates fake their responses in video interviews because job applicants' answers are a mix of honest and deceptive IMs [4]. Our models can infer the extent of job candidates' honest and deceptive IMs with various IM content domains according to all the interview questions and answers.

Fourth, self-reporting is commonly regarded as a reliable method for measuring IMs [69]. It is important to note that IMs do not always objectively reflect the subjects' emotions and mental states when assessed using sensing technology [70], [71]. Although the accuracy of assessing the ground truth in deception detection is debatable and the self-report method is susceptible to response bias [72], [73], we believe this approach is more valid for labeling deception in employment selection interviews than using observers' ratings. This is because only job candidates truly know the extent of their engagement in IMs. Moreover, well-trained observers or human interviewers have limitations in their ability to identify job candidates' IMs, as supported by empirical studies [4], [25].

Fifth, the identification of facial features relevant to truth and lie detection tasks in high-stakes contexts remains a subject of debate. Our study utilized AI techniques to analyze non-human coded features, such as temporal facial expressions and head movements, achieving an explanatory power of 79-92% in an AVI setting for distinguishing between honest and deceptive IMs. However, it is crucial to inform applicants beforehand when using AI for IM detection. Predictive results should be used to guide further follow-up interview questions, not as definitive conclusions, to adhere to ethical AI practices and legal standards in various regions.

Finally, our study might be the first to compare AI and human interviewers in detecting IMs performed by job candidates, particularly in a real hiring context. Our findings indicate that AI significantly outperforms human interviewers in this task. Although human interviewers may utilize both verbal and nonverbal cues for IM detection, AI's ability to accurately identify IMs using solely facial and head movements has proven to be substantially more effective than human judgment.

Given the growing popularity of AVIs as an interview modality [19], future research can replicate and adapt our modeling to develop additional emotion-sensing modalities for mobile and wearable devices, ensuring non-invasiveness [74]. Building on this study's utilization of the AVI platform for capturing clear facial and head imagery, subsequent research could investigate the use of face-to-face video recordings or conferencing software to expand application contexts. Additionally, future studies could incorporate other valid features from videos, such as other visual, vocal, and verbal signals, to develop multimodal, emotion-aware applications in real-life scenarios and establish their own datasets in specific contexts [24], [75].

In conclusion, our study highlights the potential of AI-powered AVIs as automated IM detectors, indicating that they could potentially serve as a copilot for human interviewers in detecting interviewees' IMs.

## Acknowledgment

The authors would like to extend special thanks for the data collection associated with this work to Tien-Hsin Hsu and Wen-Chih Lee. They also express gratitude to Ching-Fang Nien for providing the rights to use personal portraits in the published figures.

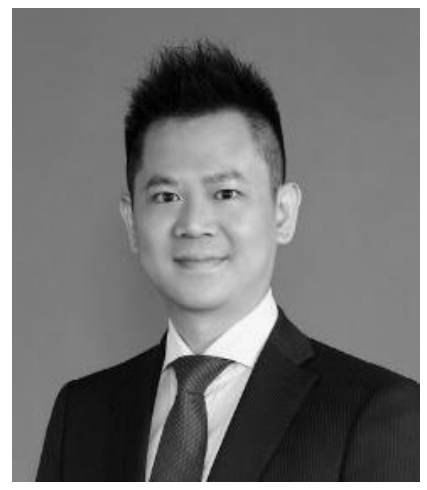

**Hung-Yue Suen** is a Distinguished Associate Professor of Technology Application and Human Resource Development at National Taiwan Normal University. His main research interests include affective computing and human–computer interaction.

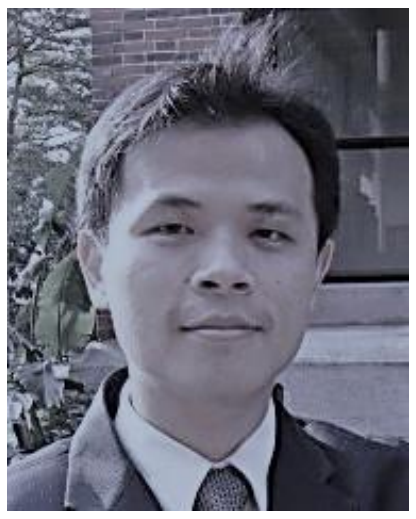

**Kuo-En Hung** is currently an industrial researcher and a Ph.D. candidate in Technology Application and Human Resource Development at National Taiwan Normal University, focusing his research on computer vision, neural networks, and human emotion simulation.

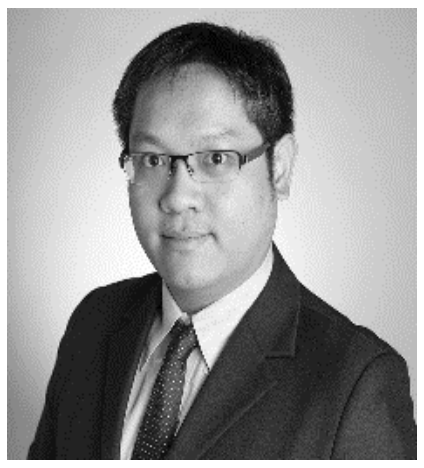

**Che-Wei Liu** serves as an Assistant Professor of Information Systems at the Kelley School of Business, Indiana University. Dr. Liu has received the Jerome Bess Faculty Fellowship. His current research focuses on Emerging Technology, as well as IT Labor and IT Management.

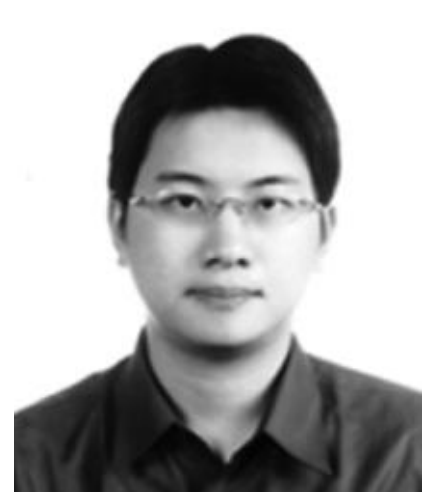

**Yu-Sheng Su** is currently an Associate Professor with the Department of Computer Science and Information Engineering, National Chung Cheng University. His research interests include cloud computing, big data analytics, intelligent system, and metaverse.

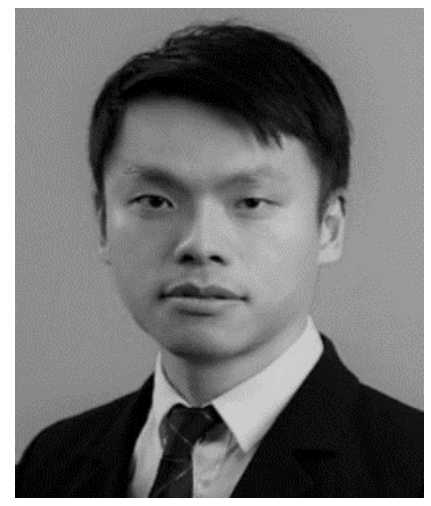

**Han-Chih Fan** is currently employed as a Human Resource Business Partner at Advanced Semiconductor Materials Lithography (ASML) in Taiwan. His research interest centers on the application of technologies in human resources.